\documentclass[mathleft]{an}

\usepackage{graphicx}
\usepackage{times}
\usepackage{natbib}
\usepackage{amssymb}

\bibpunct{(}{)}{;}{a}{}{}
\newcommand{\iaucirc}{IAU~Circ.}
\newcommand{\aap}{A\&A}
\newcommand{\aaps}{A\&AS}

\begin{document}
%%%%%%%%%%%%%%%%%%%%%%%%%%%%%%%%%%%%%%%%%%%%%%%%%%%%%%%%%%%%
\Pagespan{183}{186}
\Yearpublication{2010}
\Yearsubmission{2009}
\Month{}
\Volume{331}
\Issue{2}
\DOI{10.1002/asna.200911323}

%%%%%%%%%%%%%%%%%%%%%%%%%%%%%%%%%%%%%%%%%%%%%%%%%%%%%%%%%%%%
\title{Beginning of the Super-Soft Phase of the Classical Nova V2491 Cygni}
\author{Dai Takei\inst{1} \and Jan-Uwe Ness\inst{2}}
\titlerunning{Super-Soft X-ray Phase of V2491 Cygni}
\authorrunning{Dai Takei \& Jan-Uwe Ness}
\institute{
Department of Physics, Rikkyo University, 3-34-1 Nishi-Ikebukuro, Toshima, Tokyo 171-8501
\and
European Space Agency, XMM-Newton Observatory SOC, SRE-OAX, Apartado 78, 28691 Villanueva de la Ca\~nada, Madrid, Spain
}
%%%%%%%%%%%%%%%%%%%%%%%%%%%%%%%%%%%%%%%%%%%%%%%%%%%%%%%%%%%%
\received{2009 Dec 18}
\accepted{2009 Dec 23}
\publonline{2010 Feb 15}
%%%%%%%%%%%%%%%%%%%%%%%%%%%%%%%%%%%%%%%%%%%%%%%%%%%%%%%%%%%%
%%%%%%%%%%%%%%%%%%%%%%%%%%%%%%%%%%%%%%%%%%%%%%%%%%%%%%%%%%%%
\abstract{
 We present the results of soft X-ray studies of the classical nova V2491 Cygni using
 the Suzaku observatory. On day 29 after outburst, a soft X-ray component with a peak at
 $\sim$0.5~keV has appeared, which is tantalising evidence for the beginning
 of the super-soft X-ray emission phase. We show that an absorbed blackbody model can
 describe the observed spectra, yielding a temperature of 57~eV, neutral hydrogen column
 density of 2$\times$10$^{21}$~cm$^{-2}$, and a bolometric luminosity of
 $\sim$10$^{36}$~erg~s$^{-1}$. However, at the same time, we also found a good fit with
 an absorbed thin-thermal plasma model, yielding a temperature of 0.1~keV, neutral
 hydrogen column density of 4$\times$10$^{21}$~cm$^{-2}$, and a volume emission measure
 of $\sim$10$^{58}$~cm$^{-3}$. Owing to low spectral resolution and low signal-to-noise
 ratio below 0.6~keV, the statistical parameter uncertainties are large, but the
 ambiguity of the two very different models demonstrates that the systematic errors are
 the main point of concern. The thin-thermal plasma model implies that the soft emission
 originates from optically thin ejecta, while the blackbody model suggests that we are
 seeing optically thick emission from the white dwarf.
}
%%%%%%%%%%%%%%%%%%%%%%%%%%%%%%%%%%%%%%%%%%%%%%%%%%%%%%%%%%%%
\maketitle
%%%%%%%%%%%%%%%%%%%%%%%%%%%%%%%%%%%%%%%%%%%%%%%%%%%%%%%%%%%%

%%%%%%%%%%%%%%%%%%%%%%%%%%%%%%%%%%%%%%%%%%%%%%%%%%%%%%%%%%%%
\section{Introduction}\label{introduction}
%%%%%%%%%%%%%%%%%%%%%%%%%%%%%%%%%%%%%%%%%%%%%%%%%%%%%%%%%%%%
Classical novae are a class of cataclysmic variables, which occur in accreting binaries
with a white dwarf and a late-type companion. Hydrogen-rich material is accumulated on
the white dwarf surface. The surface temperature increases as the material accumulates.
When the accreted material reaches a critical mass for nuclear fusion, an thermonuclear
runaway process ignites on the white dwarf surface. A review of the evolution of
classical novae can be found in e.g., \cite{starrfield2008}.

After outburst, the system is expected to display X-ray emission via different
mechanisms: (1) Non-thermal emission has recently been reported from the classical nova
V24 91 Cygni by the Suzaku satellite \citep{takei2009a}. (2) Thin-thermal plasma emission
from adiabatic shocks has been reported in some classical novae, mainly in data taken
with the Swift satellite (e.g., \citealt{bode2006,ness2009a}). (3) Photospheric emission
from a hot layer of a white dwarf surface, which is characterised by bright, soft
blackbody-like continuum emission and absorption lines, similar to the class of
Super-Soft X-ray Sources (SSS; \citealt{kahabka1997}).

Following from various arguments, the time scale of SS-S emission is a function of the
white dwarf mass and the chemical composition of the post-outburst envelope (e.g.,
\citealt{sala2005,hachisu2006}). The optically thick wind theory is an approach to model
SSS light curves with some simple assumptions (e.g., \citealt{hachisu2005,hachisu2006}).
In this theory, the onset of the SSS phase is assumed to occur after the wind stops, and
the SSS phase fades out once the nuclear fuel on the white dwarf surface is
consumed. Meanwhile, recent observations have shown that significant wind velocities can
still be detected during the SSS phase (see contribution by Ness in this same journal).
It is difficult to estimate the effects of the expansion on model light curves, and,
ultimately, the white dwarf mass that would be derived from models that account for the
continued expansion during the SSS phase. More advanced models are clearly needed,
nevertheless, the duration of the SSS phase is an important observational quantity that
will always be needed to determine the white dwarf mass.

In order to determine the duration of the SSS phase, we have to estimate the turn-on and
turn-off time of SSS emission. For both quantities, dense observations in time are
needed, and intensive monitoring campaigns are currently being performed with Swift
(see, e.g., the contributions by Beardmore and by Osborne in this same journal).
However, the turn-on time is particularly difficult to be estimated because we need to
discriminate SSS emission from other sources of X-ray emission. For example, in RS\,Oph,
significant shock emission was present during the early ti-mes and faded on a time scale
of $\sim$30~days. When the SSS phase started around day 30 after outburst, some shock
em-ission was still present that may contaminate the SSS emission (e.g.,
\citealt{ness2009b}).

\cite{rohrbach2009} analysed three Chandra ACIS spectra of V1494 Aql obtained on days
134, 187, and 248 after outburst, and interpreted a soft excess in the last
observation as SSS emission. However, in the two earlier ACIS spectra, emission lines
from N were present at the same energies. They emphasised that the soft component in all
three ACIS spectra could equally well be fitted by a blackbody or a thin-thermal plasma
model with high N or O abundance. The conclusion of SSS emission in a CCD-type spectrum
has thus to be treated with care, because the energy resolution is not high enough to
compare these models. \cite{rohrbach2009} had additional high-resolution Chandra LETGS
spectra at their disposal that clearly showed that the soft component was atmospheric
rather than thin-thermal emission. For these purposes, it is thus mandatory to observe
with high energy resolution with sufficient sensitivity at soft X-ray energies so that
different models can be distinguished.

In this paper, we present the results of soft X-ray studies of the classical nova V2491
Cygni using the Suzaku observatory. Suzaku has taken two observations during the early
phase, and the X-ray CCDs aboard Suzaku provided X-ray spectra of this nova
\citep{takei2009a,takei2009b}. A tantalising hint of SSS emission was seen on day 29,
although it provides several different interpretations at the same time. We discuss
different possibilities of interpreting the soft component.

%%%%%%%%%%%%%%%%%%%%%%%%%%%%%%%%%%%%%%%%%%%%%%%%%%%%%%%%%%%%
\section{V2491 Cygni}\label{target}
%%%%%%%%%%%%%%%%%%%%%%%%%%%%%%%%%%%%%%%%%%%%%%%%%%%%%%%%%%%%
The classical nova V2491 Cygni was discovered on 2008 April 10.728~UT in the
constellation Cygnus \citep{nakano2008}. The nova was classified as an extremely fast
nova \citep{tomov2008a} from a rate of decline of $t_{2}$ $\sim$ 4.6~d
\citep{tomov2008b}, where $t_{2}$ is the time to fade by 2~mag from the optical maximum
magnitude. Based on an empirical relation between the maximum magnitude and the rate of
decline among classical novae \citep{della1995}, the distance was estimated as 10.5~kpc
\citep{helton2008}. As a consequence of the fast evolution, the white dwarf mass can be
expected to be high \citep{hachisu2009}.

Intensive monitoring observations were conducted by the Swift X-ray satellite after
outburst \citep{kuulkers2008,osborne2008,page2008,page2009}. No significant X-ray
emission was found on day 1, but it was detected on day 5. During the early phase, the
Swift spectrum exhibited extremely hard continuum emission \citep{kuulkers2008}. A
Suzaku observation was also conducted on day 9, in which the soft X-rays below
$\sim$1~keV were highly absorbed \citep{takei2009a}. We requested another Suzaku
observation to follow the evolution, and various emission lines as well as a hint of SSS
emission were found on day 29 \citep{takei2009b}. A harder spectral component is well
fitted with a thin-thermal plasma model, however, the spectra show clear signs of
attenuation in the 0.6--0.8 keV range, and the soft component cannot be fitted with the
same model as the hard part of the spectrum (see, in the bottom right panel in
figure~\ref{figure:blackbody}). The full spectral analysis of the Suzaku data is
presented by \cite{takei2009b}. The X-ray brightness increased after the Suzaku
observations, and Swift re-corded a soft and bright X-ray spectrum on day 36
\citep{osborne2008}. XMM-Newton observations were also conducted on days 40 and 50,
after the nova had entered the typical SSS phase \citep{ness2008a,ness2008b}.

%%%%%%%%%%%%%%%%%%%%%%%%%%%%%%%%%%%%%%%%%%%%%%%%%%%%%%%%%%%%
\section{Observations and reduction}\label{observation}
%%%%%%%%%%%%%%%%%%%%%%%%%%%%%%%%%%%%%%%%%%%%%%%%%%%%%%%%%%%%
V2491 Cygni was observed with Suzaku on day 29. Suzaku provides simultaneous
observations with two instruments in operation \citep{mitsuda2007}: the X-ray Imaging
Spectrometer (XIS: \citealt{koyama2007}) and the Hard X-ray Detector (HXD:
\citealt{takahashi2007,kokubun2007}), which cover the energy range of 0.2--12~keV and
10--600 keV, respectively. In order to understand the soft X-ray evolution, we
concentrate on the XIS spectrum. XIS is equipped with four X-ray CCDs at the foci of
four X-ray telescope modules \citep{serlemitsos2007}. Three of them (XIS0, 2, and 3) are
front-illuminated (FI) CCDs, sensitive in a 0.4--12~keV energy range, and the remaining
one (XIS1) is a back-illuminated (BI) CCD, sensitive at 0.2--12~keV. XIS2 has not been
functional since 2006 November, and we use the remaining CCDs. XIS was operated in
normal clocking mode with a frame time of 8~s, and the net exposure time is 25~ks. For
details of the extraction procedure, we refer to \cite{takei2009b}. In order to
understand the soft X-ray component, we fitted the respective FI and BI spectra between
0.4--0.6~keV and 0.2--0.6~keV by two different models.

%%%%%%%%%%%%%%%%%%%%%%%%%%%%%%%%%%%%%%%%%%%%%%%%%%%%%%%%%%%%
\section{Spectral Analysis}\label{analysis}
%%%%%%%%%%%%%%%%%%%%%%%%%%%%%%%%%%%%%%%%%%%%%%%%%%%%%%%%%%%%
The spectrum shows a clear detection of a soft X-ray excess below 0.6~keV
(figure~\ref{figure:blackbody}). SSS emission is the most straight forward
interpretation, however, emission lines from C, N, and O from thin-thermal plasma may
also explain the spectrum in this energy band. This is a similar case as
\cite{rohrbach2009}. A large source of uncertainty, especially in the soft band, is the
large effect of interstellar plus possibly circumstellar absorption. In the following
sections we present two model approaches to the soft spectra. The fitting was performed
with Xspec, where we used the same interstellar absorption model (TBabs module in Xspec;
\citealt{wilms2000}). Owing to the low signal-to-noise ratio, we use Cash statistics for
the determination of a best-fit model \citep{cash1979}.

\subsection{Blackbody}\label{section:bbody}

\begin{figure}[tb]
 \begin{center}
  \includegraphics[width=82mm]{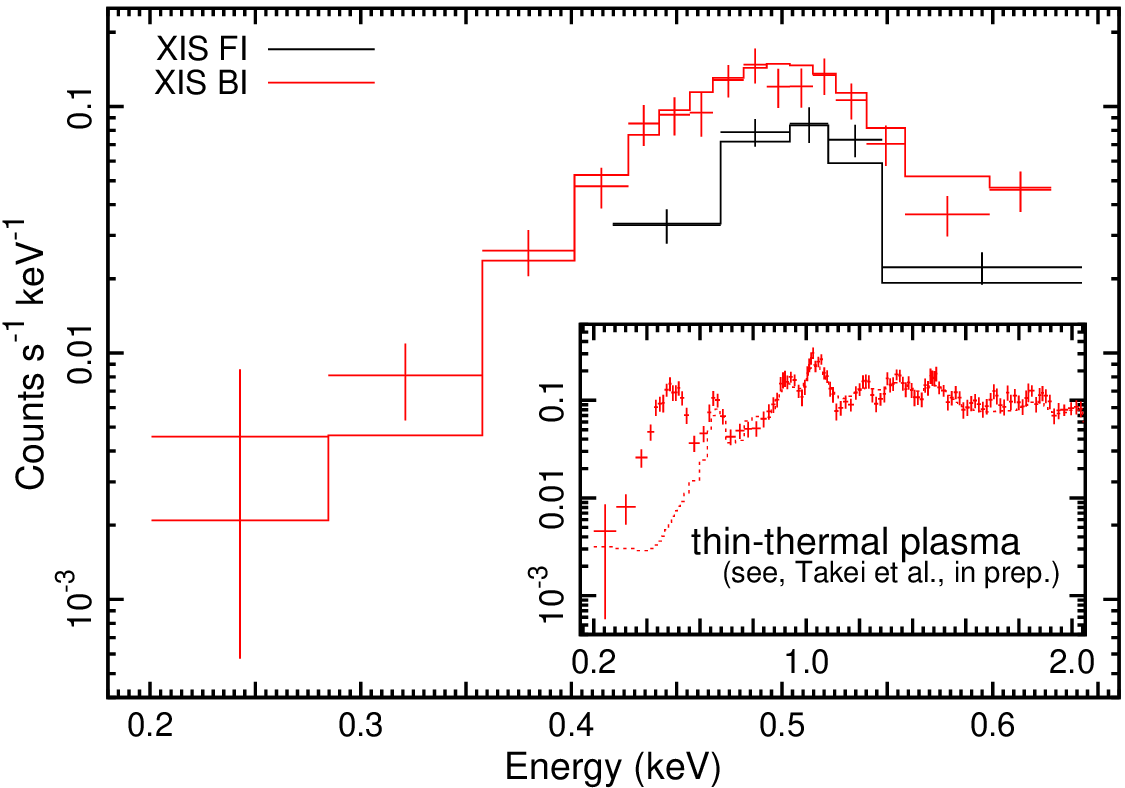}
 \end{center}
 \vspace{0mm}
 \caption{Background-subtracted XIS spectra and the best-fit blackbody model on day 29
 (see $\S$\ref{section:bbody}). The spectra are shown in pluses representing the error
 bars. The different colours represent the two instruments as indicated in the top left
 legend. The best-fit models are shown with solid lines for each instrument. The inset
 shows the hard BI spectrum and the best-fit thin-thermal plasma model presented by
 \cite{takei2009b}, illustrating the need for an additional component.
 }\label{figure:blackbody}
\end{figure}

First, we fitted the spectrum with an absorbed blackbody model. With this model we test
the possibility that the emission originates from the photosphere of the hot, extended
white dwarf. The best fit yields a blackbody temperature of 57$\pm$6~eV and an
interstellar extinction of 2$\pm$1$\times$10$^{21}$~cm$^{-2}$ with uncertainties of the
90\% confidence range. The normalisation corresponds to a bolometric luminosity of
(0.5--3)$\times$10$^{36}$~erg~s$^{-1}$ assuming a distance of 10.5~kpc by
\cite{helton2008}. The bolometric luminosity and the blackbody temperature yield a
radius of the sphere to be (470--1900)~km for the 90\% confidence level. The best-fit
model is shown in comparison to the FI and BI spectra in figure~\ref{figure:blackbody}.
The Cash goodness criterion is 17.18 with 21 bins.

\subsection{Thin-thermal plasma}\label{section:apec}

\begin{figure}[tb]
 \begin{center}
  \includegraphics[width=82mm]{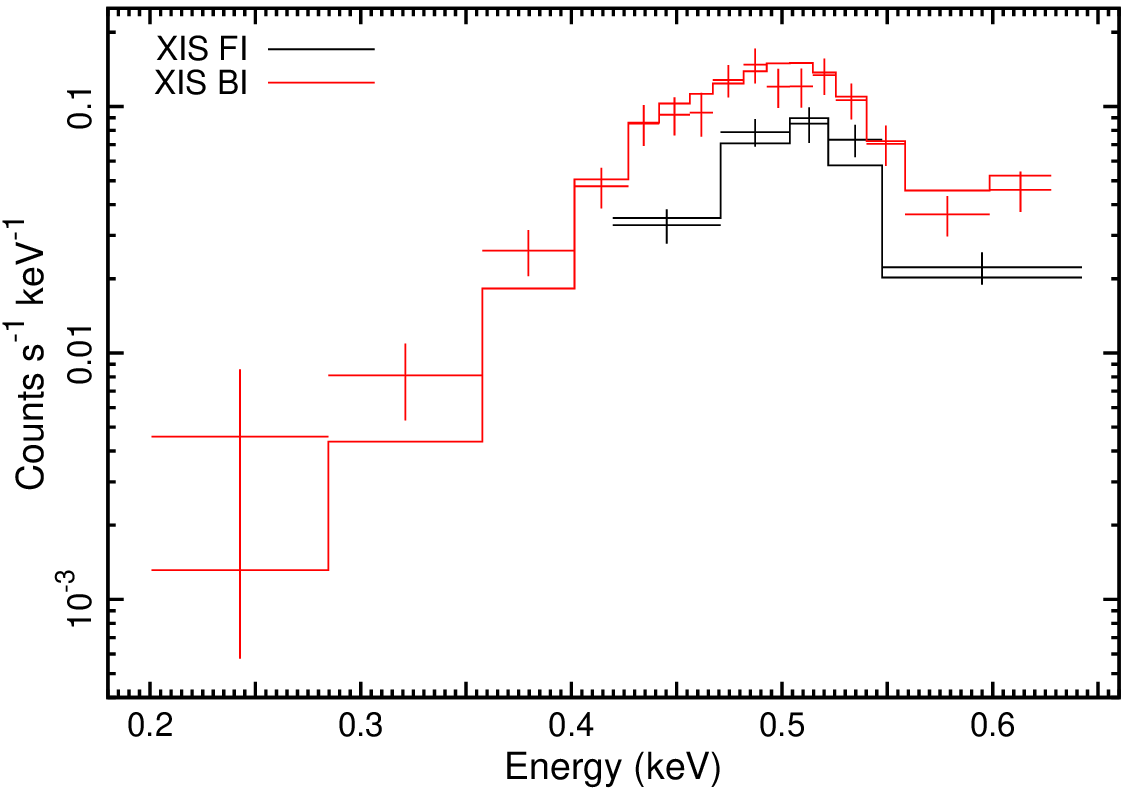}
 \end{center}
 \vspace{0mm}
 \caption{Same as Fig.~\ref{figure:blackbody} with the best-fit VAPEC model (see
 $\S$\ref{section:apec}).
 }\label{figure:apec}
\end{figure}

Next, we fitted the spectrum with a thin-thermal plasma model (VAPEC module in Xspec;
\citealt{smith2001}). This model represents optically thin plasma in collisional
equilibrium with variable abundances. We allowed the parameters temperature,
interstellar extinction, normalisation, and elemental abundances of C, N, and O to vary
as free parameters while fitting the spectrum. The other abundances were fixed at solar
abundances. The best-fit result is a temperature of $\sim$0.1~keV, an interstellar
extinction of $\sim$4$\times$10$^{21}$~cm$^{-2}$, and abundances for C, N, and O of 90,
10, and $\lesssim$6 times solar, respectively. The normalisation corresponds to a volume
emission measure of $\sim$10$^{58}$~cm$^{-3}$, assuming the distance of 10.5~kpc by
\cite{helton2008}. The best-fit model is shown in comparison to the FI and BI spectra in
figure~\ref{figure:apec}. The Cash goodness criterion is 17.30 with 21 bins.

%%%%%%%%%%%%%%%%%%%%%%%%%%%%%%%%%%%%%%%%%%%%%%%%%%%%%%%%%%%%
\section{Discussion}\label{discussion}
%%%%%%%%%%%%%%%%%%%%%%%%%%%%%%%%%%%%%%%%%%%%%%%%%%%%%%%%%%%%
The objective of this study is to understand the origin of the soft-band excess emission
peaking at $\sim$0.5~keV in a Suzaku observation taken 29 days after outburst. We have
used two different model approaches in $\S$\ref{section:bbody} and $\S$\ref{section:apec}.
Figures~\ref{figure:blackbody} and \ref{figure:apec} show that both models are good
representations of the data. While the Cash values of 17.18 and 17.30 are not a
quantitative quality criterion, their difference scales like the difference of $\chi^{2}$
(Wilks theorem; \citealt{wilks1938}). If two fits with Cash values that differ by more
than 1 (in the case of one free parameter), then the fit with the lower value is better
with a probability of 68.3\%. In our case with more than one free parameter and a
difference much less than 1, the two fits are clearly not different in their ability to
reproduce the data. The data alone are thus not sufficient to distinguish between these
two models, and both models have to be discussed.

The blackbody temperature is within the typical range of SSS spectra of classical novae,
but the bolometric luminosity appears rather low compared to the luminosities that have
been derived for other novae (e.g., \citealt{krautter1996}). Meanwhile, the luminosity
is comparable to that derived from blackbody fits to the persistent SSS CAL\,83, but the
blackbody temperature is significantly higher (e.g., \citealt{greiner1991}). This is the
reason why the derived effective radius is relatively small. While the 90\% confidence
range allows a physically realistic radius of 1900~km, it has to be noted that blackbody
fits to SSS spectra have the tendency to overestimate the bolometric luminosity and thus
the radius \citep{krautter1996}. With the small radius that we found from the blackbody
fit, the interpretation of the soft emission as originating from the extended atmosphere
around the white dwarf, thus implies a very small photospheric radius. However, it can
not be smaller than the radius of a white dwarf near the Chandrasekhar limit (e.g.,
\citealt{hamada1961}). In contrast, the emission measure of the thin-thermal plasma
model is also within the typical range (e.g., \citealt{hernanz2007,tsujimoto2009}).
Assuming that the plasma density is uniform and that the expansion velocity has not changed,
the observed value corresponds to a plasma density of $\gtrsim$10$^{6}$~cm$^{-3}$, which
is also a possible limit of novae.

An obstacle on our way to interpret the soft component as first evidence of SSS
emission, and thus the onset of the SSS phase, is the fact that a completely different
model can also explain the observations. The thin-thermal plasma model presented in
$\S$\ref{section:apec} yields an adequate fit to the data and must thus be considered
for finding the origin of the emission. A possible source for optically thin emission
are kinematic interactions of the ejecta with the surrounding medium (e.g., a planetary
nebula like in GK\,Per or the interstellar medium) or within the ejecta. In other novae,
such origins have been concluded from observations of so-called 'early hard' emission,
where the observed spectra could be fitted with APEC or MEKAL \citep{mewe1985}
models. The classical nova V2491 Cygni has been observed with extremely high velocities
(e.g., 3300~km~s$^{-1}$; \citealt{ness2008a}) and such high velocities can very well
dissipate significant amounts of energy in thin-thermal radiation. The spectral analysis
of the full Suzaku data are presented by \cite{takei2009b}, and a harder spectral
component that can only be fitted with a thin-thermal model has been detected for days
9, 29, 40, and 50. Between days 9 and 29, this component has experienced a significant
reduction in temperature. It can thus not be excluded that the soft component on day 29
has at least partly the same origin. While the soft component and the harder
thin-thermal component can not be fitted with the same model, it has to be kept in mind
that we used an isothermal model while the true plasma is likely not isothermal. With
better spectral resolution, the approach of fitting an emission measure distribution as
introduced by \cite{ness2009b} is more appropriate, and a significant amounts of cooler
plasma may be present that can account for the soft component.

%%%%%%%%%%%%%%%%%%%%%%%%%%%%%%%%%%%%%%%%%%%%%%%%%%%%%%%%%%%%
\section{Summary and Conclusions}
%%%%%%%%%%%%%%%%%%%%%%%%%%%%%%%%%%%%%%%%%%%%%%%%%%%%%%%%%%%%

The fast nova V2491 Cygni has been observed on days 9 and 29 after outburst with
Suzaku. The first observation showed a non-thermal component and an optically thin
thermal plasma component with a temperature of $\sim$2.9~keV \citep{takei2009a}. Later,
on day 29, the non-thermal component has disappeared, and the thin-thermal component has
a softer spectrum, yielding a lower temperature \citep{takei2009b}. In addition, a
super-soft component with a peak at 0.5~keV was present. The origin of this component
can either be first light of the SSS phase, or it could be the soft extension of the
thin-thermal plasma component. With the given combination of limited spectral resolution
and soft sensitivity, we can not distinguish between these two possibilities.
Technically, both scenarios are possible and plausible. Depending on the correct
interpretation, the duration of the SSS phase could be different by about a week.
According to \cite{page2009}, the SSS phase started between days 36 and 42. If the soft
component detected by Suzaku is indeed SSS emission, then the start of the SSS phase
would be at least one week earlier.

Central to addressing such questions is the spectral resolution. In high resolution, one
could clearly distinguish between SSS and a thin-thermal plasma emission. The results of
this work are one example that demonstrates the need for high spectral resolution with
sufficient sensitivity at soft energies that would have to be implemented in future
X-ray observatories like the International X-ray Observatory.

%%%%%%%%%%%%%%%%%%%%%%%%%%%%%%%%%%%%%%%%%%%%%%%%%%%%%%%%%%%%
\acknowledgements
%%%%%%%%%%%%%%%%%%%%%%%%%%%%%%%%%%%%%%%%%%%%%%%%%%%%%%%%%%%%

The authors thank the Suzaku telescope managers for allocating a part of the director's
discretionary time. D.\,T. is financially supported by the Japan Society for the
Promotion of Science. We acknowledge financial support from the Faculty of the European
Space Astronomy Centre. This research made use of data obtained from Data ARchives and
Transmission System (DARTS), provided by PLAIN center at ISAS/JAXA.

%%%%%%%%%%%%%%%%%%%%%%%%%%%%%%%%%%%%%%%%%%%%%%%%%%%%%%%%%%%%
% Bibliography
%%%%%%%%%%%%%%%%%%%%%%%%%%%%%%%%%%%%%%%%%%%%%%%%%%%%%%%%%%%%

\end{document}